\documentclass[
aps,
prd,
showpacs,
preprint,
tightenlines,
nofootinbib,
floatfix]
{revtex4}
\usepackage{graphicx,
}

\begin{document}
\title{Observables sensitive to absolute neutrino masses:\\
Constraints and correlations from world neutrino data}
\author{G.L.\ Fogli$^1$, E.\ Lisi$^1$, A.\ Marrone$^1$,
A.\ Melchiorri$^2$, A.\ Palazzo$^1$, P.\ Serra$^2$, J.\ Silk$^3$}

\address{
$^1$ Dipartimento di Fisica and Sezione INFN
di Bari, Via Amendola 173, 70126, Bari, Italy\\
$^2$ Dipartimento di Fisica and Sezione INFN, Universit\`a
degli Studi di Roma ``La Sapienza'', P.le Aldo Moro 5, 00185, Rome,
Italy\\
$^3$ Astrophysics, Denys Wilkinson Building, Keble Road, OX13RH,
Oxford, United Kingdom}


\begin{abstract}
In the context of three-flavor neutrino mixing, we present a
thorough study of the phenomenological constraints applicable to
three observables sensitive to absolute neutrino masses: The
effective neutrino mass in Tritium beta decay $(m_\beta)$; the
effective Majorana neutrino mass in neutrinoless double beta decay
$(m_{\beta\beta})$; and the sum of neutrino masses in cosmology
$(\Sigma)$. We discuss the correlations among these variables which
arise from the combination of all the available neutrino oscillation
data, in both normal and inverse neutrino mass hierarchy. We set
upper limits on $m_\beta$ by combining updated results from the
Mainz and Troitsk experiments. We also consider the latest results
on $m_{\beta\beta}$ from the Heidelberg-Moscow experiment, both with
and without the lower bound claimed by such experiment.  We derive
upper limits on $\Sigma$ from an updated combination of data from
the Wilkinson Microwave Anisotropy Probe (WMAP) satellite and the 2
degrees Fields (2dF) Galaxy Redshifts Survey, with and without
Lyman-$\alpha$ forest data from the Sloan Digital Sky Survey (SDSS),
in models with a non-zero running of the spectral index of
primordial inflationary perturbations. The results are discussed in
terms of two-dimensional projections of the globally allowed region
in the $(m_\beta,m_{\beta\beta},\Sigma)$ parameter space, which
neatly show the relative impact of each data set. In particular, the
(in)compatibility between $\Sigma$ and $m_{\beta\beta}$ constraints
is highlighted for various combinations of data.  We also briefly
discuss how future neutrino data (both oscillatory and
non-oscillatory) can further probe the currently allowed regions.
\end{abstract}
 \maketitle


\section{introduction}

Atmospheric \cite{Atmo,SKLE,MACR,Soud}, solar
\cite{Home,SAGE,GALL,Catt,SKso,SNOs}, reactor \cite{KamL,Grat,KL04}
and accelerator \cite{K2K1,K2K2} neutrino experiments have
convincingly established that neutrinos are massive and mixed
\cite{Pont,Maki}. World neutrino data (with a single controversial
exception \cite{LSND}) are consistent with a three-flavor mixing
framework \cite{Conc,Barg,Keow,Ours,Malt,Srub}, parameterized in
terms of three neutrino masses $(m_1,m_2,m_3)$ and of three mixing
angles $(\theta_{12},\theta_{23},\theta_{13})$ \cite{PDG4}, plus a
possible CP violating phase $\delta$ \cite{Kays}.

Neutrino oscillation experiments are sensitive to two independent
squared mass difference, $\delta m^2$ and $\Delta m^2$ (with $\delta
m^2\ll \Delta m^2$), hereafter defined as \cite{Delt}
\begin{equation}
\label{DeltaDef} (m^2_1,m^2_2,m^2_3)= \mu^2 + \left( -\frac{\delta
m^2}{2}, +\frac{\delta m^2}{2},\pm\Delta m^2 \right),
\end{equation}
where $\mu$ fixes the absolute neutrino mass scale, while the cases
$+\Delta m^2$ and $-\Delta m^2$ identify the so-called normal and
inverted neutrino mass hierarchies, respectively. Neutrino
oscillation data indicate that $\delta m^2\simeq 8\times 10^{-5}$
eV$^2$ \cite{KL04} and $\Delta m^2\simeq 2.4\times 10^{-3}$ eV$^2$
\cite{SKLE,K2K2}. They also indicate that $\sin^2\theta_{12}\simeq
0.3$ \cite{KL04}, $\sin^2\theta_{23}\simeq 0.5$ \cite{SKLE,K2K2},
and $\sin^2\theta_{13}\lesssim \mathrm{few}\%$ \cite{CHOO}. However,
they are currently unable to determine the mass hierarchy
($\pm\Delta m^2$) and the phase $\delta$ \cite{Kays}, and are
insensitive to the absolute mass parameter $\mu$ in
Eq.~(\ref{DeltaDef}).

The absolute neutrino mass scale can be probed by non-oscillatory
neutrino experiments. The most sensitive laboratory experiments to
date have been focussed on tritium beta decay \cite{Holz,Wein,Mass}
and on neutrinoless double beta decay \cite{Doub,Vo02,Elli}. Beta
decay experiments probe the so-called effective electron neutrino
mass $m_\beta$ \cite{Mbet},
\begin{equation}
\label{mb} m_\beta =
\left[c^2_{13}c^2_{12}m^2_1+c^2_{13}s^2_{12}m^2_2+s^2_{13}m^2_3
\right]^\frac{1}{2}\ ,
\end{equation}
where $c^2_{ij}=\cos^2\theta_{ij}$ and $s^2_{ij}=\sin^2\theta_{ij}$.
Current experiments (Mainz \cite{Main} and Troitsk \cite{Troi})
provide upper limits in the range $m_\beta\lesssim \mathrm{few}$~eV
\cite{PDG4,Eite}.

Neutrinoless double beta decay ($0\nu2\beta$) experiments are
instead sensitive to the so-called effective Majorana mass
$m_{\beta\beta}$ (if neutrinos are Majorana fermions)
\cite{Wo81,BiPe},
\begin{equation}\label{mbb}
m_{\beta\beta} = \left|
c^2_{13}c^2_{12}m_1+c^2_{13}s^2_{12}m_2e^{i\phi_2}+s^2_{13}m_3
e^{i\phi_3}\right|\ ,
\end{equation}
where $\phi_2$ and $\phi_3$ parameterize relative (and unknown)
Majorana neutrino phases \cite{ScVa}. All $0\nu2\beta$ experiments
place only upper bounds on $m_{\beta\beta}$ (the most sensitive
being in the eV range \cite{Elli}), with the exception of the
Heidelberg-Moscow experiment \cite{Kl01}, which claims a positive
(but highly debated \cite{Elli}) $0\nu2\beta$ signal, corresponding
to $m_{\beta\beta}$ in the sub-eV range at best fit
\cite{Kl03,Kl04}.

Recently, astrophysical and cosmological observations have started
to provide indirect upper limits on absolute neutrino masses (see,
e.g., the reviews \cite{Barg,Mass,Dolg}), competitive with those
from laboratory experiments. In particular, the combined analysis of
high-precision data from Cosmic Microwave Background (CMB)
anisotropies and Large Scale Structures (LSS) has already reached a
sensitivity of $O(\mathrm{eV})$ (see, e.g., \cite{Be03,Tg04,Laha})
for the sum of the neutrino masses $\Sigma$,
\begin{equation}\label{Sigma}
\Sigma = m_1+m_2+m_3\ .
\end{equation}
We recall that the total neutrino energy density in our Universe,
$\Omega_{\nu}h^2$ (where $h$ is the Hubble constant normalized to
$H_0=100$ km~s$^{-1}$~Mpc$^{-1}$) is related to $\Sigma$ by the
well-known relation $\Omega_{\nu}h^2=\Sigma / (93.2 \mathrm{\ eV})$
\cite{PDG4}, and plays an essential role in theories of structure
formation. It can thus leave key signatures in LSS data (see, eg.,
\cite{Hu98}) and, to a lesser extent, in CMB data (see, e.g.,
\cite{Ma95}). Very recently, it has also been shown that accurate
Lyman-$\alpha$ (Ly$\alpha$) forest data \cite{Mc04}, taken at face
value, can improve the current CMB+LSS constraints on $\Sigma$ by a
factor of $\sim 3$, with important consequences on absolute neutrino
mass scenarios \cite{Se04}. Given the potentially large systematics
of Ly$\alpha$ data \cite{Mc04}, we will conservatively perform and
compare global cosmological fits both with and without Ly$\alpha$
constraints.

There is a vast literature about the theoretical and
phenomenological links and correlations connecting  [via
Eqs.~(\ref{DeltaDef})--(\ref{Sigma})] the $3\nu$ mass-mixing
oscillation parameters $(\delta m^2,\Delta m^2,\theta_{ij})$ with
the observables sensitive to absolute neutrino masses
$(m_\beta,m_{\beta\beta},\Sigma)$; see, e.g., the extensive
bibliography in the reviews \cite{Barg,Keow,Mass,Elli}. These
interesting correlations have often been studied and plotted in
terms of an auxiliary variable, such as the lightest neutrino mass
$m_\mathrm{lightest}$; in particular, many relevant papers have 
focussed on
constraints in the $(m_{\beta\beta},m_\mathrm{lightest})$ plane
(see, e.g., \cite{Ligh,Simk}). However, since $m_\mathrm{lightest}$
is not directly measured, it can actually be removed from data
analyses, by charting the results just in terms of the three
measurable quantities $(m_\beta,m_{\beta\beta},\Sigma)$.%
\footnote{Similar considerations apply to the auxiliary variable
$\mu$ in Eq.~(\ref{DeltaDef}), being
$m^2_\mathrm{lightest}=\mu^2-\delta m^2/2$ for normal hierarchy, and
$m^2_\mathrm{lightest}=\mu^2-\Delta m^2$ for inverted hierarchy.}

In this work we undertake a global phenomenological analysis of the
constraints applicable to the observables
$(m_\beta,m_{\beta\beta},\Sigma)$, by using up-to-date experimental
data and state-of-the-art calculations for all the relevant
laboratory and astrophysical quantities. To our knowledge, this is
the first attempt to fitting world neutrino data, both oscillatory
and non-oscillatory, in the ``standard'' astroparticle physics
scenario emerged in the last few years (i.e., standard $3\nu$ mixing
plus standard cosmology). The results are shown and discussed in
terms of allowed regions in each of the three 2-dimensional
projections of the parameter space
$(m_\beta,m_{\beta\beta},\Sigma)$, for both normal and inverted
hierarchy. Such projections neatly show the interplay among
different observables, as well as the impact of each data set. In
particular, a tension is seen to emerge between current constraints
on $\Sigma$ and $m_{\beta\beta}$. Our synthesis of world neutrino
data appears also useful for estimating the sensitivity required by
future experiments in order to solve, e.g., the mass hierarchy
ambiguity, or to probe Majorana phases.

The structure of the paper is as follows. In Secs.~II, III, IV, and
V we describe our treatment of the relevant input from oscillation,
beta-decay, $0\nu2\beta$, and cosmological data, respectively. In
Sec.~VI we discuss the bounds on $(m_\beta,m_{\beta\beta},\Sigma)$
coming from neutrino oscillation data only, with emphasis on the
correlations among couples of observables. In Sec.~VII we combine
the previous results with the upper bounds on
$(m_\beta,m_{\beta\beta},\Sigma)$ coming from laboratory experiments
and cosmological CMB+LSS data (without Ly$\alpha$ forest data). In
Sec.~VIII we implement a possible lower bound on $m_{\beta\beta}$
based on the claim in \cite{Kl03,Kl04}, and describe its effect on
the global fit. Finally, in Sec.~IX we discuss the impact of recent
Ly$\alpha$ data \cite{Mc04,Se04}. We draw our conclusions in Sec.~X.

\section{Input from $\nu$ oscillation data}

In this work, we perform an updated global analysis of oscillation
neutrino data including: Atmospheric Super-Kamiokande (SK) data
\cite{SKLE}; solar data from the Chlorine (Cl) \cite{Home}, Gallium
(Ga) \cite{SAGE,GALL,Catt}, SK \cite{SKso}, and Sudbury Neutrino
Observatory (SNO) \cite{SNOs} experiments; reactor data from the
KamLAND \cite{KamL,KL04} and CHOOZ \cite{CHOO} experiments; and
KEK-to-Kamioka (K2K) accelerator data \cite{K2K1,K2K2}. We refer the
reader to  previous works \cite{Prev} for details of the analysis.
We include the latest experimental data presented at the Neutrino
2004 Conference \cite{Catt,Grat,K2K2}. In particular, we make a
$3\nu$ analysis of the available solar and KamLAND data, which are
sensitive to the mass-mixing parameters $(\delta
m^2,\sin^2\theta_{12},\sin^2\theta_{13})$. We derive the likelihood
of the $\Delta m^2$ parameter from a graphical reduction and
combination of the corresponding likelihoods determined by the SK
atmospheric \cite{SKLE} and K2K accelerator \cite{K2K2} experiments
in the $2\nu$
limit,%
\footnote{At present, a full $3\nu$ analysis of SK+K2K data could
only be jointly performed by the two experimental collaborations,
since given relevant information is either unpublished or not
reproducible outside the collaborations. In any case, $3\nu$ effects
are known to produce only negligible or minor changes in the $\Delta
m^2$ likelihood from SK+K2K data (within current bounds on
$\theta_{13}$), and thus they are not expected to alter the results
of this work in any significant way.}
marginalized with respect to $\theta_{23}$.%
\footnote{The angle $\theta_{23}$ is not relevant for the
calculation of the $(m_\beta,m_{\beta\beta},\Sigma)$ observables,
see Eqs.~(\ref{mb})--(\ref{Sigma}).}
The probability distribution of $\Delta m^2$ is crucial to obtain
upper bounds on $\theta_{13}$ from the CHOOZ reactor experiment
\cite{Th13}. The results of our analysis of neutrino oscillation
data are embedded in a $\chi^2$ fit function,
\begin{eqnarray}\label{Chi2osc}
\lefteqn{\chi^2_\mathrm{osc}
(\delta m^2,\Delta m^2,\sin^2\theta_{12},\sin^2\theta_{13})=}
\nonumber\\
&&\chi^2_\mathrm{solar+KamL}(\delta
m^2,\sin^2\theta_{12},\sin^2\theta_{13})+\nonumber\\
&&\chi^2_\mathrm{SK+K2K}(\Delta m^2)
+\chi^2_\mathrm{CHOOZ}(\Delta m^2,\sin^2\theta_{13})\ ,
\end{eqnarray}
which provides the likelihood for the four oscillation parameters
$(\delta m^2,\Delta m^2,\sin^2\theta_{12},\sin^2\theta_{13})$ needed
for the calculation of the $(m_\beta,m_{\beta\beta},\Sigma)$
observables.

Figure~1 shows the function
$\Delta\chi^2_\mathrm{osc}=\chi^2_\mathrm{osc}-\min\chi^2_\mathrm{osc}$
from the global $\nu$ oscillation data fit, marginalized with
respect to each of its four arguments. The condition
$\Delta\chi_\mathrm{osc}^2=n^2$ provides $n$-$\sigma$ bounds on each
mass-mixing parameter. At best fit we find
\begin{eqnarray}
\sqrt{\delta m^2} &\simeq &9.1 \times 10^{-3}\mathrm{\ eV}\ ,\\
\sqrt{\Delta m^2} &\simeq &4.9 \times 10^{-2}\mathrm{\ eV}\ ,
\end{eqnarray}
and
\begin{eqnarray}
\sin^2\theta_{12} &\simeq& 0.29\ ,\\
\sin^2\theta_{13} &\simeq & 0.007\ .
\end{eqnarray}
Notice, however, that the slight preference for nonzero
$\sin^2\theta_{13}$ in Fig.~1 is not statistically significant.

\section{Input from Tritium $\beta$-decay data}

Updated determinations of the effective electron neutrino mass
squared $m^2_\beta$ have been recently presented \cite{Eite} for the
Mainz \cite{Main} and Troitsk \cite{Troi} tritium $\beta$-decay
experiments. The experimental values are consistent with zero within
errors:
\begin{eqnarray}
m^2_\beta&=&-1.2\pm3.0 \mathrm{\ eV}^2\ (\mathrm{Mainz})\ ,\label{Mainz}\\
m^2_\beta&=&-2.3\pm3.2 \mathrm{\ eV}^2\ (\mathrm{Troitsk})\
,\label{Troitsk}
\end{eqnarray}
where errors are at $1\sigma$ level, and systematic and statistical
components have been added in quadrature. The $\chi^2_\beta$
function associated to the above $\beta$-decay data can be simply
defined as:
\begin{equation}
\label{chi2b} \chi^2_{\beta,X}(m_\beta)=\left(
\frac{m^2_\beta-m^2_{\beta,X}} {\sigma_{\beta,X}}\right)^2 \ ,
\end{equation}
where $m^2_{\beta,X}$ and $\sigma_{\beta,X}$ are the central values
and errors on the right hand side of Eqs.~(\ref{Mainz}) and
(\ref{Troitsk}), for $X=\mathrm{Mainz}$ and $X=\mathrm{Troitsk}$,
respectively. By restricting the domain to the physical region
($m^2_\beta\geq 0$), the $\Delta\chi^2$ function relevant for our
analysis becomes
\begin{equation}\label{DeltachibX}
\Delta\chi^2_{\beta,X}(m_\beta) =
\chi^2_{\beta,X}(m_\beta)-\chi^2_{\beta,X}(0)\ ,
\end{equation}
providing the following upper bounds at 95\% C.L.\
($\Delta\chi^2\simeq 4$),
\begin{eqnarray}
m_\beta&<&2.2 \mathrm{\ eV}\ (\mathrm{Mainz})\ ,\label{UMainz}\\
m_\beta&<&2.1 \mathrm{\ eV}\ (\mathrm{Troitsk})\ ,\label{UTroitsk}
\end{eqnarray}
in agreement with the upper bounds quoted by the experimental
collaborations \cite{Eite}.

It is matter of debate whether the two experiments have common
systematics, e.g., responsible for the negative central value of
$m^2_\beta$ (see \cite{Brad} for a recent discussion). In the
absence of a clear indication in this sense, we assume that the two
experimental results are independent, and simply combine them
through
\begin{equation}\label{Deltachib}
\Delta\chi^2_{\beta}(m_\beta) =\Delta\chi^2_{\beta,\mathrm{Mainz}}
(m_\beta)+\Delta\chi^2_{\beta,\mathrm{Troitsk}}(m_\beta)\ .
\end{equation}
From the above definition, we get a combined upper limit at 95\%
C.L.,
\begin{equation}\label{MainzTroitsk}
m_\beta<1.8~\mathrm{eV}\ (\mathrm{Mainz}+\mathrm{Troitsk})\ ,
\end{equation}
which is less conservative than the $3$~eV upper limit recommended
in \cite{PDG4}. However, as we will see, upper limits on $m_\beta$
in the 2--3 eV range are, in any case, too weak to contribute
significantly to the current global fit in the
$(m_\beta,m_{\beta\beta},\Sigma)$ parameter space, so that
``conservativeness'' is not (yet) an issue in this context.

\section{Input from Germanium $0\nu2\beta$ decay data}

Neutrinoless double beta decay processes of the kind
$(Z,A)\to(Z+2,A)+2e^-$ have been searched in many experiments with
different isotopes, yielding negative results (see \cite{Vo02,Elli}
for reviews). Recently, members of the Heidelberg-Moscow experiment
\cite{HMex} have claimed the detection of a $0\nu2\beta$ signal from
the $^{76}$Ge isotope \cite{Kl01,Kl03,Kl04}. The claimed signal
would correspond to a decay half-life $T$ in years (y) within the
following $3\sigma$ interval \cite{Kl03,Kl04}:
\begin{equation}
\label{T} \log_{10}(T/\mathrm{y})=25.08^{+0.54}_{-0.24}\ (3\sigma)\
.
\end{equation}

If the $0\nu2\beta$ signal is entirely due to light Majorana
neutrino masses, the half-life $T$ is related to the
$m_{\beta\beta}$ parameter in Eq.~(\ref{mbb}) by the relation
\begin{equation}
\label{Cdef} m^2_{\beta\beta}=\frac{m^2_e}{C_{mm}T}\ ,
\end{equation}
where $m_e$ is the electron mass and $C_{mm}$ is the nuclear matrix
element for the considered isotope \cite{Elli}.

Unfortunately, theoretical uncertainties on $C_{mm}$ are rather
large (see e.g.\ \cite{Elli}), and their---somewhat
arbitrary---estimate is matter of debate (see \cite{Simk,Civi,Bahc}
and refs.\ therein). Our approach to estimate a central value and an
error for matrix element $C_{mm}$ relevant to $^{76}$Ge experiments
is the following. We take from the list of ``acceptable'' $C_{mm}$
calculations (i.e., calculations with no subsequently recognized
errors or biases) discussed in \cite{Elli} the ``extremal''
estimates: $\log_{10}(C_{mm}^\mathrm{min}/\mathrm{y}^{-1})=-13.85$
\cite{Mini} and
$\log_{10}(C_{mm}^\mathrm{max}/\mathrm{y}^{-1})=-12.88$ \cite{Maxi}.
Then we assume that their half-sum and their difference define,
respectively, the central value and the $3\sigma$ uncertainty (in
log scale), namely: $\log_{10}(C_{mm}^0/\mathrm{y}^{-1})=-13.36$ and
$3\sigma\equiv 13.85-12.88=0.97$. We thus take
\begin{equation}\label{Cval}
\log_{10} (C_{mm}/\mathrm{y}^{-1}) = -13.36\pm 0.97\ (3\sigma)
\end{equation}
which embeds a rather conservative error estimate---the $3\sigma$
uncertainty being almost equal to an order of magnitude variation in
$C_{mm}$, as also evaluated in \cite{Bahc}. The above central value
for $C_{mm}$ is close to recent detailed theoretical calculations
\cite{Rodi}.

From the previous equations, under the assumption of a positive
$0\nu2\beta$ signal in Ref.~\cite{Kl03,Kl04}, we derive that
\begin{eqnarray}
\log_{10}(m_{\beta\beta}/\mathrm{eV})&=&\log_{10}(m_e/\mathrm{eV})
-\frac{1}{2}\log_{10}(C_{mm}/\mathrm{y}^{-1})-
\frac{1}{2}\log_{10}(T/\mathrm{y})
\label{logmbb1}\\
&=& 5.71 +\frac{1}{2} (13.36\pm0.97)
-\frac{1}{2}\left(25.08^{+0.54}_{-0.24} \right)\
\label{logmbb2}\\
&=& -0.23\pm 0.53 \ (3\sigma)\label{logmbb3}\ ,
\end{eqnarray}
where, in the last line, we have symmetrized the asymmetric range
$25.08^{+0.54}_{-0.24}$ as $25.23\pm0.39$ (according to the recipe
suggested in \cite{Dago}) before adding the errors in quadrature
(see \cite{Dago} for the statistical rationale of this procedure).
Equation~(\ref{logmbb3}) allows us to define the $\Delta
\chi^2_{\beta\beta}(m_{\beta\beta})$ function associated to the
claim in Ref.~\cite{Kl03,Kl04}, and provides the $3\sigma$ range
$0.17<m_{\beta\beta}<2.0$ (in eV).
\footnote{Our estimated $3\sigma$ range $m_{\beta\beta}\in
[0.17,2.0]$ eV overlaps---but does not coincide---with the range
$m_{\beta\beta}\in[0.1,0.9]$ eV quoted in \cite{Kl03,Kl04}, since we
use different estimates for the central value and uncertainties of
$C_{mm}$.}

The claim in \cite{Kl03,Kl04} has been subject to strong criticism,
especially after the first publication \cite{Kl01} (see \cite{Elli}
and refs.\ therein). Therefore, we will also consider the
possibility that $T=\infty$ is allowed (i.e., that there is no
$0\nu2\beta$ signal), in which case the experimental lower bound on
$m_{\beta\beta}$ disappears. Since one cannot really reprocess the
data of \cite{Kl03,Kl04} under the assumption of no signal, nor
combine such data with the negative results of other experiments
with comparable sensitivity to $m_{\beta\beta}$ \cite{Elli}, the
definition of a $\Delta \chi^2_{\beta\beta}(m_{\beta\beta})$
function in the absence of a $0\nu2\beta$ signal is somewhat
ambiguous and, in part, arbitrary. Our approach simply consists in
stretching the lower error in Eq.~(\ref{logmbb3}) to infinity, so
that the lower bound on $m_{\beta\beta}$ disappears, while the upper
bound remains in the $O(\mathrm{eV})$ ballpark indicated by the most
sensitive experiments to date \cite{Elli}. It is worth noticing that
none of the main results of our work depends on the precise
definition of the upper bound on $m_{\beta\beta}$.

In conclusion, we adopt the following two possible $0\nu2\beta$
inputs (and associated $\Delta\chi^2_{\beta\beta}$ functions) for
our global analysis:
\begin{eqnarray}
\log_{10}(m_{\beta\beta}/\mathrm{eV})&=&-0.23\pm0.18\ (0\nu2\beta
\mathrm{\ signal\ assumed})\ , \label{bbinput1}\\
\log_{10}(m_{\beta\beta}/\mathrm{eV})&=&-0.23^{+0.18}_{-\infty}\
(0\nu2\beta \mathrm{\ signal\ not\ assumed})\ , \label{bbinput2}
\end{eqnarray}
where errors are at $1\sigma$ level. Concerning the unknown Majorana
phases $\phi_2$ and $\phi_3$ in Eq.~(\ref{mbb}), we simply assume
that they are independent and uniformly distributed in the range
$[0,\pi]$.

\section{Input from cosmological data}

The neutrino contribution to the overall energy density of the
universe can play a relevant role in large scale structure formation
and leave key signatures in several cosmological data sets. More
specifically, neutrinos suppress the growth of fluctuations on
scales below the horizon when they become non relativistic. A
massive neutrinos of a fraction of eV would therefore produce a
significant suppression in the clustering on small cosmological
scales (namely, for comoving wavenumber $k\sim 0.05 \ h\
\mathrm{Mpc}^{-1}$).

To constrain $\Sigma$ from cosmological data, we perform a
likelihood analysis comparing the recent observations with a set of
models with cosmological parameters sampled as follows: cold dark
matter (cdm) density $\Omega_\mathrm{cdm}h^2 \in [0.05,0.20]$ in
steps of $0.01$; baryon density $\Omega_{b}h^2 \in [0.015, 0.030]$
(motivated by Big Bang Nucleosynthesis) in steps of $0.001$; a
cosmological constant $\Omega_{\Lambda} \in [0.50, 0.96]$ in steps
of  $0.02$; and neutrino density $\Omega_{\nu}h^2 \in [0.001,
0.020]$ in steps of $0.002$. We restrict our analysis to {\it flat}
$\Lambda$-CDM models, $\Omega_\mathrm{tot}=1$, and we add a
conservative external prior on the age of the universe, $t_0 > 10$
Gyrs. The use of a method based on a database of models instead of
Markov Chains (see e.g. \cite{Be03}) has the advantage of being more
reliable in the definition of likelihood confidence contours at more
than 3$\sigma$ level. No alternative model to a cosmological
constant is considered: the dark energy is described by an
unclustered fluid with constant equation of state ($w=-1$).
Variations in the equation of state affect mainly curvature
parameters (see, e.g., \cite{bean}) and therefore do not alter our
results on $\Sigma$.

The value of the Hubble constant in our database is not an
independent parameter, since it is determined through the flatness
condition. We adopt the conservative top-hat bound $0.50 < h < 0.90$
and we also consider the $1\sigma$ constraint on the Hubble
parameter, $h=0.71\pm0.07$, obtained from Hubble Space Telescope
(HST) measurements~\cite{freedman}. We allow for a reionization of
the intergalactic medium by varying the CMB photon optical depth
$\tau_c$ in the range $\tau_c \in [0.05,0.30]$ in steps of $0.02$.

We restrict the analysis to adiabatic inflationary models with a
negligible contribution of gravity waves. We do not consider the
possibility of isocurvature perturbations \cite{jxd} or topological
defects (see, e.g., \cite{dkm}). We let vary the spectral index $n$
of scalar primordial fluctuations in the range $n\in [0.85, 1.3]$
and its running $dn/d\ln k \in [-0.40,0.2]$ assuming pivot scales at
$k_0=0.05 \mathrm{\ Mpc}^{-1}$ and $k_0=0.002 \mathrm{\ Mpc}^{-1}$.
We rescale the fluctuation amplitude by a prefactor $C_{110}$, in
units of the value $C_{110}^\mathrm{WMAP}$ measured by the Wilkinsin
Microwave Anisotropy Probe (WMAP) satellite. Finally, concerning the
neutrino parameters, we fix the number of neutrino species to
$N_{\nu}=3$, all with the same mass (the effect of mass differences
compatible with neutrino oscillation being negligible in the current
cosmological data \cite{pastor}). An higher number of neutrino
species can weakly affect both CMB and LSS data (see, e.g.,
\cite{bowen}) but is highly constrained by standard big bang
nucleosynthesis and is not considered in this work, where we focus
on $3\nu$ mixing.


The cosmological data we considered comes from observation of CMB
anisotropies and polarization, galaxy redshift surveys and
luminosity distances of type Ia supernovae. For the CMB data we use
the recent temperature and cross polarization results from the WMAP
satellite \cite{Be03} using the method explained in \cite{map5} and
the publicly available code on the LAMBDA web site \cite{LAMBDA}.
Given a theoretical temperature anisotropy and polarization angular
power spectrum in our database, we can therefore associate a
$\chi^2_\mathrm{WMAP}$ to the corresponding theoretical model.

We further include the latest results from the
BOOMERanG-98~\cite{ruhl}, Degree Angular Scale Interferometer
(DASI)~\cite{halverson}, MAXIMA-1~\cite{lee}, Cosmic Background
Imager (CBI)~\cite{pearson}, and Very Small Array Extended
(VSAE)~\cite{clive} experiments by using the publicly available
correlation matrices, window functions and beam and absolute
calibration errors. The CMB data analysis methods for the non-WMAP
data have been already extensively described in various papers (see,
e.g., \cite{mesilk}) and will not be reported here.

In addition to the CMB data we also consider the real-space power
spectrum of galaxies from either the 2 degrees Fields (2dF) Galaxy
Redshifts Survey or the Sloan Digital Sky Survey (SDSS), using the
data and window functions of the analysis of \cite{thx} and
\cite{Tg04}. We restrict the analysis to a range of scales over
which the fluctuations are assumed to be in the linear regime ($k <
0.2 h^{-1}\mathrm{\ Mpc}$). When combining with the CMB data, we
marginalize over a bias $b$ for each data set considered as an
additional free parameter.

We also include information from the Ly$\alpha$ Forest in the SDSS,
using the results of the analysis of \cite{Se04} and \cite{Mc04},
which probe the amplitude of linear fluctuations at very small
scales. For this data set, small-scale power spectra are computed at
high redshifts and compared with the values presented in
\cite{Mc04}. As in \cite{Se04}, we do not consider running.

We finally incorporate constraints obtained from the SN-Ia
luminosity measurements of \cite{riess} using the so-called GOLD
data set. Luminosity distances at SN-Ia redshifts are computed for
each model in our database and compared with the observed apparent
bolometric SN-Ia luminosities.


In Fig.~2 we plot the likelihood distribution for $\Sigma$ from our
joint analysis of CMB~+~SN-Ia~+~HST~+~LSS data, transformed into an
equivalent $\Delta\chi^2_\Sigma$ function, which allows to derive
bounds on $\Sigma$ at any fixed confidence level. We take LSS data
either from the SDSS or the 2dF survey (dashed and solid curves,
respectively).%
\footnote{For the sake of brevity, the subdominant block of data
(SN-Ia~+~HST) is not explicitly indicated in figure labels.}

As we can see, these curves do not show evidence for a neutrino mass
(the best fit being at $\Sigma\simeq 0$) and provide the $2\sigma$
bound $\Sigma \lesssim 1.4$ eV. Such bound is in good agreement with
previous results in similar analyses
\cite{Be03,hannestad,Tg04,barger, crotty}. However our results can
be considered complementary and, in some cases, independent from
those analyses. For example, the recent SN-Ia data from \cite{riess}
were not included in those analyses, so that a model with
$\Omega_m=1$ and break in the spectrum as in \cite{sarkar} could
produce a good fit to the CMB and LSS data, strongly increasing the
upper limits on $\Sigma$. Models of this kind are ruled out by the
inclusion of the SN-Ia data.

We found that the CMB+LSS results in Fig.~2 are stable under the
assumption of running, i.e. there is a weak correlation between the
running and the neutrino masses in the range of scales probed by
these data sets. For definiteness, we will use the combination
including 2dF data (solid line in Fig.~2) in the global analyses of
Secs.~VII and VIII.

Also plotted in Fig.~2 is the  $\Delta\chi^2_\Sigma$ function from a
joint analysis of CMB~+~SN-Ia~+~HST~+~2dF~+~Ly$\alpha$. No running
is assumed in this analysis, and we find a $2\sigma$ bound $\Sigma <
0.47$ eV, in very good agreement (despite the more approximate
method we used) with the analysis already presented in \cite{Se04}.

As shown in Fig.~2 and already discussed in \cite{Se04}, the
inclusion of the Ly$\alpha$ data from the SDSS set greatly improves
the constraints on $\Sigma$. However, we remark that the constraints
on the linear density fluctuations obtained from this dataset are
derived from measurement of the Ly$\alpha$ flux power spectrum
$P_F(k)$ after a long inversion process, which involves numerical
simulations and marginalization over the several parameters of the
Ly$\alpha$ model. Since the effects of possible systematics need
still to be explored further both observationally and theoretically,
we here take a conservative approach, and discuss the implications
of this results for neutrino physics in a separate section
(Sec.~IX).

\section{Constraints on $(m_\beta,m_{\beta\beta},\Sigma)$
from oscillation data only}

In this Section we present and discuss the regions allowed by
neutrino oscillation data in the parameter space
$(m_\beta,m_{\beta\beta},\Sigma)$. Figure~3  shows the projections
of the $2\sigma$ regions ($\Delta\chi^2_\mathrm{osc}$=4), onto each
of the three coordinate planes $(m_{\beta\beta},\Sigma)$,
($m_{\beta\beta}$,$m_\beta$), and ($m_\beta,\Sigma$), for both
normal hierarchy (thick solid curves) and inverted hierarchy (thin
solid curves). In each of the three coordinate planes, the
observables appear to be strongly correlated. Such correlations can
be largely understood in the approximation $\sin^2\theta_{13}\simeq
0$, and by distinguishing the following three main cases for
Eq.~(\ref{DeltaDef}): a) $\mu^2 \gg\Delta m^2$;  b) $\mu^2 \gtrsim
\Delta m^2$; and c) $\mu^2<\Delta m^2$. Earlier discussions of
correlations in the $(m_{\beta\beta},\Sigma)$  and
($m_{\beta\beta}$,$m_\beta$) planes have been presented in
\cite{Whis,Glas} and in \cite{Mats}, respectively.

\subsection{Degenerate spectrum (DS)}

For $\mu^2\gg\Delta m^2$ in Eq.~(\ref{DeltaDef}), the neutrino
masses $m_i$ ($i=1,2,3$) form a degenerate spectrum (DS) and, in
both hierarchies, for $s^2_{13}\simeq 0$ one has
\begin{eqnarray}
m_i &\simeq& (\mu,\mu,\mu)\ , \\
m_\beta &\simeq& \mu\ , \\
m_{\beta\beta} &\simeq & \mu\, |c^2_{12}+s^2_{12}e^{i\phi_2}|\ ,\\
\Sigma &\simeq & 3\mu \ .
\end{eqnarray}
By eliminating the auxiliary mass scale parameter $\mu$, one gets
the following linear correlations among observables:
\begin{eqnarray}
m_\beta &\simeq& \frac{\Sigma}{3}, \label{QD1}\\
m_{\beta\beta} &\simeq & \frac{\Sigma}{3}\, f\ ,\label{QD2}\\
m_{\beta\beta} &\simeq & m_\beta \, f \ ,\label{QD3}
\end{eqnarray}
where the factor
\begin{equation}\label{effe}
f=|c^2_{12}+s^2_{12}e^{i\phi_2}|\ ,
\end{equation}
which can take any value in the range $c^2_{12}-s^2_{12}\leq f\leq
1$, embeds our ignorance about the Majorana phase $\phi_2$.

Equation~(\ref{QD1}) explains the tight correlation in the
$(m_\beta,\Sigma)$ plane of Fig.~3, where, for DS masses, the
allowed region reduces to a ``line.'' The correlation is instead
relaxed by the variable factor $f$ in the $(m_{\beta\beta},\Sigma)$
and $(m_{\beta\beta},m_\beta)$ planes, where the allowed regions
appear as ``strips'' in the DS limit.

\subsection{Partially degenerate (PD) spectrum}

For $\mu^2\gtrsim \Delta m^2$ in Eq.~(\ref{DeltaDef}), the neutrino
mass spectrum can be considered as partially degenerate (PD), in the
sense that the largest (smallest) mass splitting can(not) be
resolved. For vanishing $s^2_{13}$ and $\delta m^2$ one has
\begin{eqnarray}
m_i &\simeq& (\mu,\mu,\sqrt{\mu^2\pm\Delta m^2})\ , \\
m_\beta &\simeq& \mu\ , \\
m_{\beta\beta} &\simeq & \mu\, |c^2_{12}+s^2_{12}e^{i\phi_2}|\ ,\\
\Sigma &\simeq & 2\mu + \sqrt{\mu^2\pm\Delta m^2}\ ,
\end{eqnarray}
where the upper (lower) sign refers to normal (inverted) hierarchy.
By eliminating $\mu$, the following correlations are obtained:
\begin{eqnarray}
\Sigma &\simeq& 2 m_\beta + \sqrt{m^2_\beta\pm\Delta m^2} \ , \label{PD1}\\
m_{\beta\beta} &\simeq & m_\beta\, f\ ,\label{PD2}\\
\Sigma & \simeq & \frac{2 m_{\beta\beta} + \sqrt{m^2_{\beta\beta}\pm
f\Delta m^2}}{f}\ ,\label{PD3}
\end{eqnarray}
where $f$ is defined as in Eq.~(\ref{effe}).

According to Eq.~(\ref{PD1}), the regions allowed in the
$(m_\beta,\Sigma)$ plane of Fig.~3  for normal and inverted
hierarchies---which overlap in the degenerate spectrum case---branch
out when the spectrum becomes partially degenerate and sensitive to
$\pm \Delta m^2$. In particular, the curve for inverted hierarchy
bends upwards and ends at
$(m_{\beta,\mathrm{min}},\Sigma_\mathrm{min})\simeq (\sqrt{\Delta
m^2},2\sqrt{\Delta m^2})$, while the curve for normal hierarchy
bends downwards, and eventually enters the regime of hierarchical
spectrum discussed in the next section. The two hierarchies are
instead not distinguishable in the $(m_{\beta\beta},m_\beta)$ plane
of Fig.~3, where Eq.~(\ref{PD2}) provides, in the PD case, the same
hierarchy-independent linear correlation as in the DS case
[Eq.~(\ref{QD3})].

Finally, Eq.~(\ref{PD3}) explains the correlation in the plane
($m_{\beta\beta},\Sigma$); in particular, by taking extremal values
of $f$ (1 and $c^2_{12}-s^2_{12}$) at fixed $m_{\beta\beta}$, one
gets upper and lower bounds on $\Sigma$ in the PD case, which are
different in the normal hierarchy ($+\Delta m^2$) and inverted
hierarchy ($-\Delta m^2$) cases \cite{Whis,Glas}; analogously, for
fixed $\Sigma$ one gets upper and lower bounds on $m_{\beta\beta}$.
In conclusion, it appears that for PD spectra the cases of normal
and inverted hierarchy might be discriminated (the better the lower
the neutrino masses) in the planes $(m_\beta,\Sigma)$ and
($m_{\beta\beta},\Sigma$), but not in the plane
$(m_{\beta\beta},m_\beta)$.

\subsection{Hierarchical spectrum (HS)}

Finally, only for {\em normal\/} mass hierarchy one can also
consider cases with $\mu^2<\Delta m^2$, where the smallest mass
splitting cannot be neglected, and a truly ``hierarchical spectrum''
(HS) with three resolved masses ($m_1<m_2\ll m_3$) is obtained. For
$s^2_{13}\simeq 0$ one gets in the HS case:
\begin{eqnarray}
m_i &\simeq& \left(\sqrt{\mu^2-\frac{\delta
m^2}{2}},\sqrt{\mu^2+\frac{\delta m^2}{2}},
\sqrt{\mu^2+\Delta m^2}\right)\ , \label{SH1}\\
m_\beta &\simeq& \sqrt{\mu^2-\delta m^2(c^2_{12}-s^2_{12})/2}\ , \label{SH2}\\
m_{\beta\beta} &\simeq & \left|c^2_{12}\sqrt{\mu^2-\frac{\delta
m^2}{2}}+s^2_{12}e^{i\phi_2}
\sqrt{\mu^2+\frac{\delta m^2}{2}}\right|\ ,\label{SH3}\\
\Sigma &\simeq & \sqrt{\mu^2-\frac{\delta
m^2}{2}}+\sqrt{\mu^2+\frac{\delta m^2}{2}}+\sqrt{\mu^2+\Delta m^2}\
.\label{SH4}
\end{eqnarray}
Eliminating the auxiliary mass scale $\mu$ from the above equations
does not lead to particular transparent formulae, and we shall limit
ourselves to a few comments. For $\mu^2=\delta m^2/2$ one gets from
the above equations the minima $m_{\beta,\mathrm{min}}\simeq
\sqrt{s^2_{12}\delta m^2}$ and $\Sigma_\mathrm{min}\simeq
\sqrt{\delta m^2}+\sqrt{\Delta m^2+\delta m^2/2}$, but not the
minimum of $m_{\beta\beta}$. In fact,
$m_{\beta\beta,\mathrm{min}}\simeq 0$ is reached at $e^{i\phi_2}=-1$
and for values of $\mu^2$ slightly higher than $\delta m^2/2$ in
Eq.~(\ref{SH3}), which also lead to values of $m_\beta$ and $\Sigma$
above their minima. Therefore, while the allowed region for normal
hierarchy has an endpoint at
$(m_{\beta,\mathrm{min}},\Sigma_\mathrm{min})$ in the
$(m_\beta,\Sigma)$ plane, it continues indefinitely  ($\log
m_{\beta\beta}\to-\infty$) in the other two planes. We also mention
that, in the HS case, corrections induced by nonzero values of
$s^2_{13}$ (not included in the above equation, but included in the
fit of Fig.~3) can be nonnegligible, and contribute to the spread of
the allowed regions for small values of the three parameters
$(m_\beta,m_{\beta\beta},\Sigma)$.

We conclude this section with a few remarks on prospective
improvements. Besides current neutrino oscillation experiments, a
number of new experiments have been planned for the next decade,
aiming at a better determination of the mass-mixing neutrino
parameters \cite{Futu}, and in particular of $\theta_{13}$
\cite{Fu13}, which is currently unknown. A nonzero value of
$\theta_{13}$ is crucial to discriminate normal and inverted
hierarchies $(\pm \Delta m^2)$ through matter effects in future
long-baseline experiments \cite{NuFa}. If the neutrino oscillation
parameters were all perfectly known, the allowed regions in the
upper panel of Fig.~3 would shrink to two lines (for the two
possible hierarchies). In the other two panels the allowed regions
would be only moderately reduced, given the irreducible spread in
$m_{\beta\beta}$ (for any fixed value of $m_\beta$ or $\Sigma$)
induced by the unknown Majorana phases, and in particular by the
phase $\phi_2$ through the factor $f$ in Eq.~(\ref{effe}).

\section{Adding upper bounds to $(m_\beta,m_{\beta\beta},\Sigma)$ from
laboratory and cosmology}

The results of the previous sections have been obtained by
projecting $\Delta\chi^2=4$ regions from neutrino oscillation data
only ($\Delta \chi^2=\Delta\chi^2_\mathrm{osc}$). Here we consider
also the regions obtained by projecting the various $\Delta\chi^2$
functions associated with the $m_\beta$ data input (Sec.~III), with
the $m_{\beta\beta}$ input in the conservative case [Sec.~IV,
Eq.~(\ref{bbinput2})] and with the cosmological CMB+LSS input
(Sec.~V), both separately and in combination with
$\Delta\chi^2_\mathrm{osc}$.%
\footnote{We will add Ly$\alpha$ data to the global fit in Sec.~IX.}

Figure~4 shows such projections in the three coordinate planes.
Separate laboratory and cosmological upper bounds at the $2\sigma$
level are shown as dashed lines, while the regions allowed by the
combination of laboratory, cosmological, and oscillation data are
shown as thick solid curves for normal hierarchy and as thin solid
curves for inverted hierarchy. It can be seen that the upper bounds
on the $(m_\beta,m_{\beta\beta},\Sigma)$ observables are dominated
by the cosmological upper bound on $\Sigma$. This bound, via the
$(m_\beta,\Sigma)$ and $(m_{\beta\beta},\Sigma)$ correlations
induced by oscillation data, provides upper limits also on
$m_{\beta\beta}$ and $m_\beta$, which happen to be stronger than the
current laboratory limits by a factor $\sim 4$. Since significant
improvements on laboratory limits for $m_{\beta\beta}$ and $m_\beta$
will require new experiments and several years of data taking
\cite{Eite,Avig}, cosmological determinations of $\Sigma$, although
indirect, will continue to provide, in the next future, the most
sensitive upper limits (and hopefully a signal) for absolute
neutrino mass observables.

Figure~4 is useful to estimate the prospective impact of possible
future measurements of the kind $M\pm \sigma_M$, where $M$ is any
one of the three observables  $(m_\beta,m_{\beta\beta},\Sigma)$. The
intersections of the currently allowed regions with the prospective
$M\pm \sigma_M$ band(s) provide immediately an estimate of
``future'' allowed regions in the presence of one or more positive
signals. In particular, we invite the reader to evaluate by herself
or himself the relative accuracy needed to discriminate---if
possible at all---the two mass spectrum hierarchies in the three
planes of Fig.~4, as well as the $m_{\beta\beta}$ accuracy needed to
reduce the vertical spread (largely due to the Majorana phase factor
$f$) of the allowed bands in the two lower panels. It will then be
clear that probing the spectrum hierarchy and the Majorana phase(s)
with future experiments will be a very challenging task.

\section{Adding lower bounds on $m_{\beta\beta}$ from the claimed $0\nu2\beta$ signal}

In this Section we keep the same data set as in Sec.~VII, except for
the $0\nu2\beta$ data input, which we now take from
Eq.~(\ref{bbinput1}), in order to show the effect of the signal
claimed in \cite{Kl03,Kl04} on the global fit.

Figure~5 shows that, in this case, there is a lower bound on
$m_{\beta\beta}$ at $2\sigma$, as indicated by an additional
horizontal dashed line (not present in the previous Fig.~4). This
lower bound is somewhat ``too high'' to allow a good combined fit
with oscillation and cosmological CMB+LSS data (beta decay limits
still being not relevant in the global fit). Therefore, the global
$2\sigma$ allowed region extends somewhat outside the $2\sigma$
limits from cosmology and $0\nu2\beta$ data taken separately --- a
clear indication of some statistical tension between such data.

From Fig.~5 it follows that, assuming standard $3\nu$ mixing {\em
and\/} the $0\nu2\beta$ claim in \cite{Kl03,Kl04}: 1) The preferred
spectrum is degenerate and there is no possibility to discriminate
normal and inverted hierarchy in the
$(m_\beta,m_{\beta\beta},\Sigma)$ parameter space; 2) A significant
fraction of the $2\sigma$ region allowed by the current global fit
is within the sensitivity reach ($m_\beta\sim 0.2$--0.3 eV) of the
future beta-decay experiment KATRIN \cite{Eite}; 3) A cosmological
signal for $\Sigma\sim O(1)$~eV should be ``around the corner;'' 4)
if such a signal will not be found, the tension between $0\nu2\beta$
and $\Sigma$ data will increase, unless the theoretical central
value for the matrix element $C_{mm}$ is significantly shifted
upwards, so as to decrease the preferred values of $m_{\beta\beta}$.
Indeed, we shall see in the next section that such tension is
already increased by including recent Ly$\alpha$ data.

Let us speculate, however, about the possibility that the {\em
true\/} values of $(m_\beta,m_{\beta\beta},\Sigma)$ lye within the
allowed regions of Fig.~5. It appears that, in order to
significantly reduce such regions, one should: improve the
sensitivity to $m_\beta$ by a factor $\sim 5$ at least, find a
cosmological signal of $\Sigma\sim O(\mathrm{eV})$ with an
uncertainty definitely smaller than a factor $\sim 2$, and confirm
the current $0\nu2\beta$ signal claim with a total
(experimental+theoretical) error reduced by better than a factor of
$2$ (with respect to our estimate). Improvements in neutrino
oscillation data would instead produce marginal effects (not shown).
These stringent requirements set the stage for possible future tests
of the absolute neutrino mass scenario allowed by the global data
fit in Fig.~5.

\section{Adding L\lowercase{y}$\alpha$ forest data}

In this Section we re-evaluate the bounds shown in Sec.~VII (without
$0\nu2\beta$ signal) and in Sec.~VIII (with $0\nu2\beta$ signal), by
adding the recent Ly$\alpha$ forest data \cite{Mc04,Se04} discussed
in Sec.~V.

\subsection{Case with no $0\nu2\beta$ signal}

Figure~6 shows the global fit assuming no $0\nu2\beta$ signal, but
adding Ly$\alpha$ data. The same fit, without Ly$\alpha$ data, has
previously shown in Fig.~4. By comparing Figs.~4 and 6 the impact of
Ly$\alpha$ data, taken at face value, is impressive: The upper limit
on $\Sigma$ is improved by a factor $\sim 3$ and, through the
correlations induced by neutrino oscillation data constraints, it is
transformed into upper limits onto $m_\beta$ and $m_{\beta\beta}$,
which are an order of magnitude stronger than the current laboratory
ones. The overall bounds are strong enough to approach the regime of
partially degenerate spectrum, where the bands for normal and
inverted hierarchies start to branch out in the two left panels of
Fig.~6.

The bounds in Fig.~6 set very stringent requirements to future
laboratory experiments sensitive to absolute neutrino mass. A factor
of $\gtrsim 10$ improvement is required both for $m_\beta$ and for
$m_{\beta\beta}$ measurements, in order to be competitive with the
cosmological upper bounds and to improve current upper limits in the
$(m_\beta,m_{\beta\beta},\Sigma)$ parameter space; finding a signal
would be even more demanding. On the other hand, cosmological data
should eventually provide evidence for nonzero $\Sigma$ in the range
$0.05\lesssim\Sigma/\mathrm{eV}\lesssim 0.5$ for normal hierarchy,
or in the range $0.09\lesssim \Sigma/\mathrm{eV}\lesssim 0.5$ for
inverted hierarchy.

\subsection{Case with claimed $0\nu2\beta$ signal}

The strong upper bound on $\Sigma$ obtained by adding Ly$\alpha$
data to the cosmological data input increases the already existing
tension with the $0\nu2\beta$ claim (see Sec.~VIII) to the point
that a global fit would provide meaningless and unstable results.

For this reason, we show in Figure~7 the bands separately allowed at
$2\sigma$ by the $0\nu2\beta$ claim (horizontal band) and by the
combination of all oscillation and cosmological (CMB+2dF+Ly$\alpha$)
neutrino data (lower slanted bands), for both normal and inverted
mass hierarchy. Only the plane $(m_{\beta\beta},\Sigma)$ is shown in
Fig.~7, since current laboratory bounds on $m_\beta$ are an order of
magnitude away from the global allowed region, as previously
discussed.

In Fig.~7, the absence of overlap between the bands separately
allowed at $2\sigma$ is a clear symptom of possible problems, either
in some data sets or in their theoretical interpretation, which
definitely prevent any global combination of data. However, it would
be premature to conclude that, e.g., the $0\nu2\beta$ claim is
``ruled out'' by cosmological data. Firstly, cosmological bounds on
$\Sigma$ are rather indirect, and involve the processing of a large
amount of data, whose systematics must be more closely scrutinized,
in particular for the latest Ly$\alpha$ data set. Secondly, one
cannot exclude that future calculations of the nuclear matrix
element $C_{mm}$ in $0\nu2\beta$ may be revised upwards, thus
lowering the $m_{\beta\beta}$ allowed range and relaxing the tension
with current upper limits on $\Sigma$. Thirdly, one cannot exclude
that $0\nu2\beta$ decay might receive contributions from new physics
effects beyond the exchange of light Majorana neutrinos. Finally,
either some data or some fundamental assumptions about the standard
three-neutrino mixing and the cosmological scenarios might be wrong.
With all the above cautionary remarks, it is anyway exciting that
global neutrino data analyses have already reached a point where
such fundamental questions start to arise.

\section{Conclusions and perspectives}

In the context of standard $3\nu$ mixing and standard cosmology, we
have performed a global phenomenological analysis of the constraints
applicable to three observables sensitive to absolute neutrino
masses: the effective neutrino mass in Tritium beta decay
$(m_\beta)$; the effective Majorana neutrino mass in neutrinoless
double beta decay $(m_{\beta\beta})$; and the sum of neutrino masses
in cosmology $(\Sigma)$.

We have first discussed the correlations among such variables
induced by neutrino oscillation data, in both normal and inverse
neutrino mass hierarchy (see Fig.~3 and related comments). We have
then applied laboratory constraints on $m_\beta$ and
$m_{\beta\beta}$, as well as cosmological constraints on $\Sigma$,
in order to obtain global fits in the
$(m_\beta,m_{\beta\beta},\Sigma)$ parameter space, which embed all
world neutrino data relevant to absolute neutrino mass scenarios.
The results have been discussed in terms of two-dimensional
projections of the globally allowed region in the
$(m_\beta,m_{\beta\beta},\Sigma)$ parameter space, which neatly show
the relative impact of each data set. In particular, the
(in)compatibility between $\Sigma$ and $m_{\beta\beta}$ constraints
has been discussed for various combinations of data, as shown in
Figs.~4 and 5 (without and with the $0\nu2\beta$ signal claim,
respectively) and in Figs.~6 and 7 (which  include recent Ly$\alpha$
data).

We have also briefly discussed how future neutrino data (both
oscillatory and non-oscillatory) with improved sensitivity can
further probe the currently allowed regions. Our graphical
representations appear to be rather useful in this sense, since
prospective measurements of any of the three observable
$(m_\beta,m_{\beta\beta},\Sigma)$ can be easily mapped onto the
currently allowed regions.

\acknowledgments

The work of G.L.F., E.L., A.M., and A.P.\ is supported by the
Italian Ministero dell'Istruzione, Universit\`a e Ricerca (MIUR) and
Istituto Nazionale di Fisica Nucleare (INFN) through the
``Astroparticle Physics'' research project. G.L.F., E.L., and A.M.
would like to thank the Organizers of the {\em Neutrino 2004\/}
Conference (whose scientific results have been largely used in this
work) for kind hospitality in Paris. We thank D.\ Montanino,  S.\
Pastor, S.T.\ Petcov, W.\ Rodejohan, and S.\ Sarkar for useful comments.


\newpage
\begin{figure}
\vspace*{-0cm}\hspace*{-2.5cm}
\includegraphics[scale=0.9, bb= 30 100 500 700]{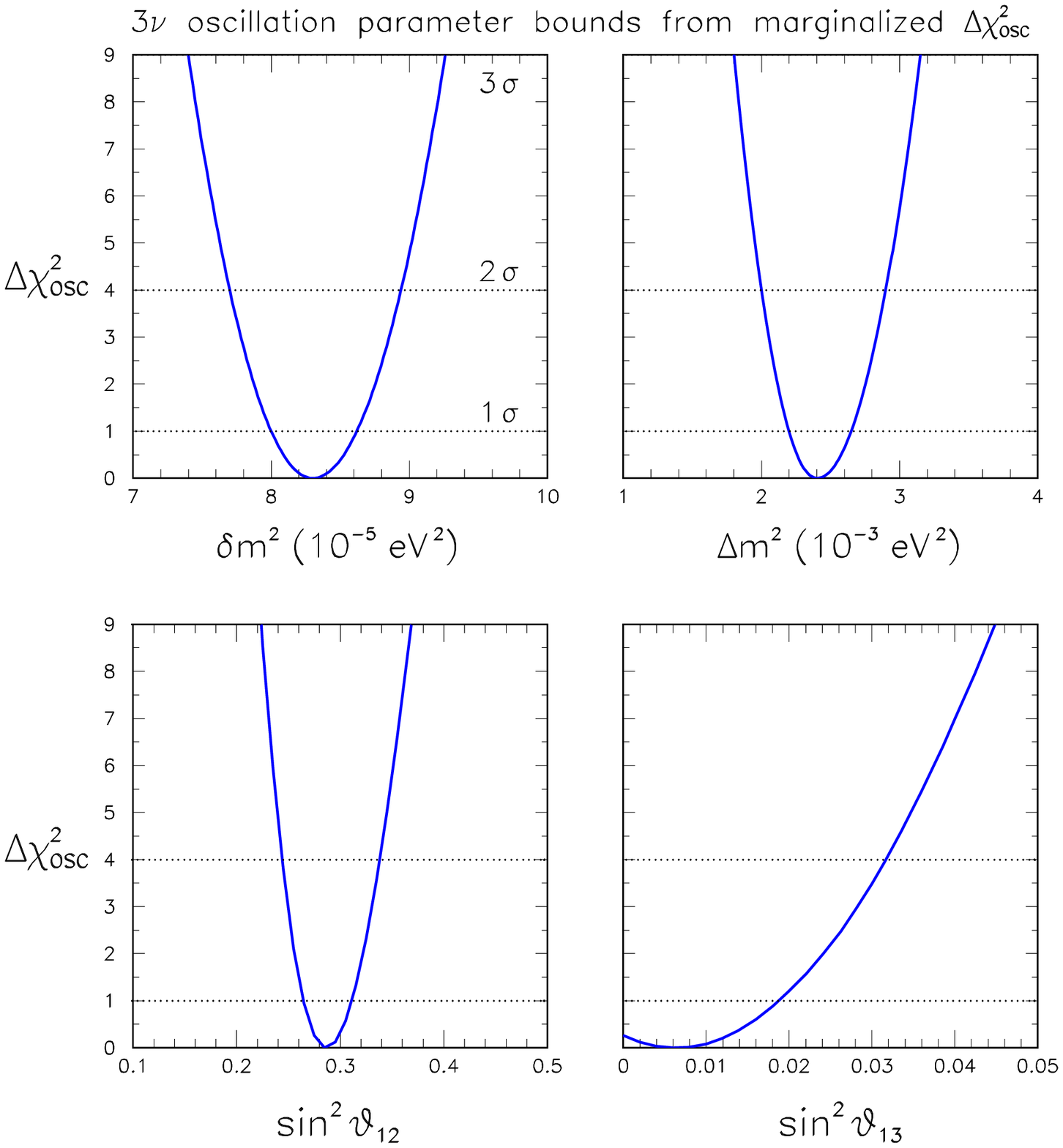}
\vspace*{-1cm} \caption{\label{fig1} Bounds on the $3\nu$
mass-mixing neutrino parameters $(\delta m^2,\Delta
m^2,\sin^2\theta_{12},\sin^2\theta_{13})$ from our analysis of
current neutrino oscillation data. The solid curves represent
projections of the $\Delta\chi^2_\mathrm{osc}$ fit function. The
intercepts of the curves with the horizontal dotted lines at
$\Delta\chi^2=n^2$ define the $n$-$\sigma$ limits on each parameter.
There is no statistically significant lower limit for
$\sin^2\theta_{13}$.}
\end{figure}

\begin{figure}
\vspace*{-0cm}\hspace*{-2.5cm}
\includegraphics[scale=0.9, bb= 30 100 500 700]{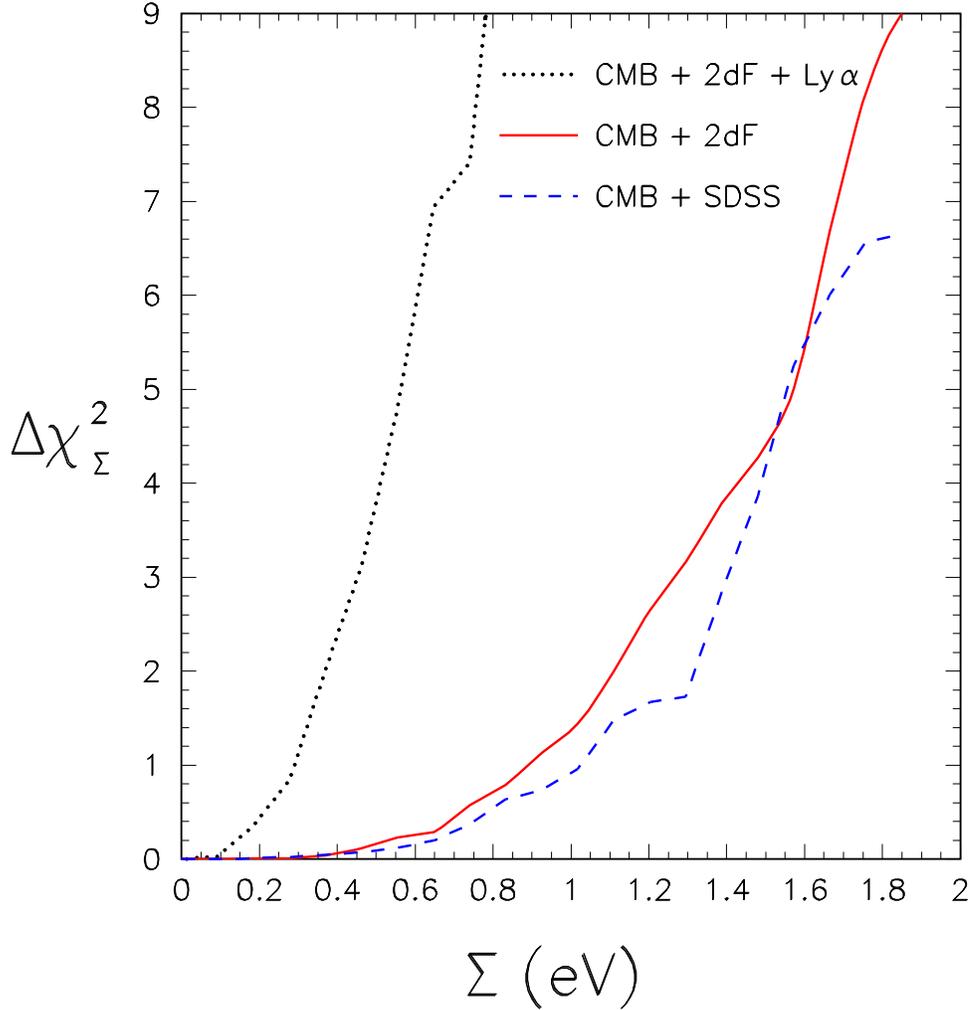}
\vspace*{-1cm} \caption{\label{fig2} Upper bounds on the sum of
neutrino masses $\Sigma$ from our $3\nu$ analysis of cosmological
data, given in terms of the $\Delta\chi^2_\Sigma$ function. The
solid and dashed curves refer to the combination of  CMB and LSS
data (CMB+2dF and CMB+SDSS, respectively). The two CMB+LSS fits
provide comparable results and, for definiteness, the CMB+2df one is
adopted. In addition, we consider also the case where the recent
Ly$\alpha$ data from the SDSS are included, providing significantly
stronger constraints on $\Sigma$ (dotted curve). See the text for
details.}
\end{figure}

\begin{figure}
\vspace*{-0cm}\hspace*{-2.5cm}
\includegraphics[scale=0.9, bb= 30 100 500 700]{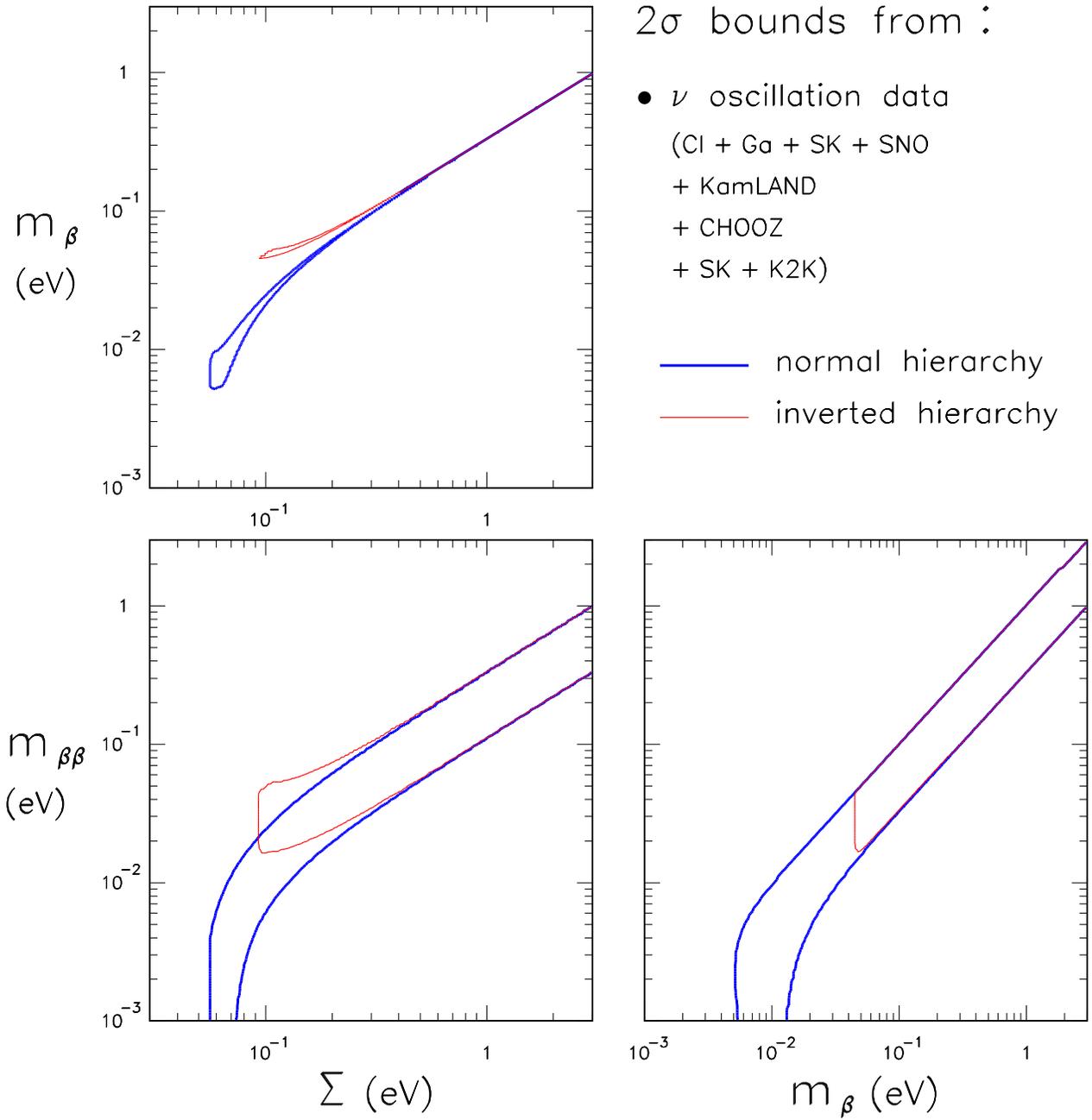}
\vspace*{-1cm} \caption{\label{fig3} Global $3\nu$ analysis in the
$(m_{\beta},m_{\beta\beta},\Sigma)$ parameter space, using
oscillation data only. The three panels show the two-dimensional
projections of the volume allowed at $\Delta\chi^2_\mathrm{osc}=4$
($2\sigma$ on each parameter) for both normal hierarchy (thick solid
curves) and inverted hierarchy (thin solid curves), respectively.
The emerging correlations between parameters are discussed in the
text.}
\end{figure}

\begin{figure}
\vspace*{-0cm}\hspace*{-2.5cm}
\includegraphics[scale=0.9, bb= 30 100 500 700]{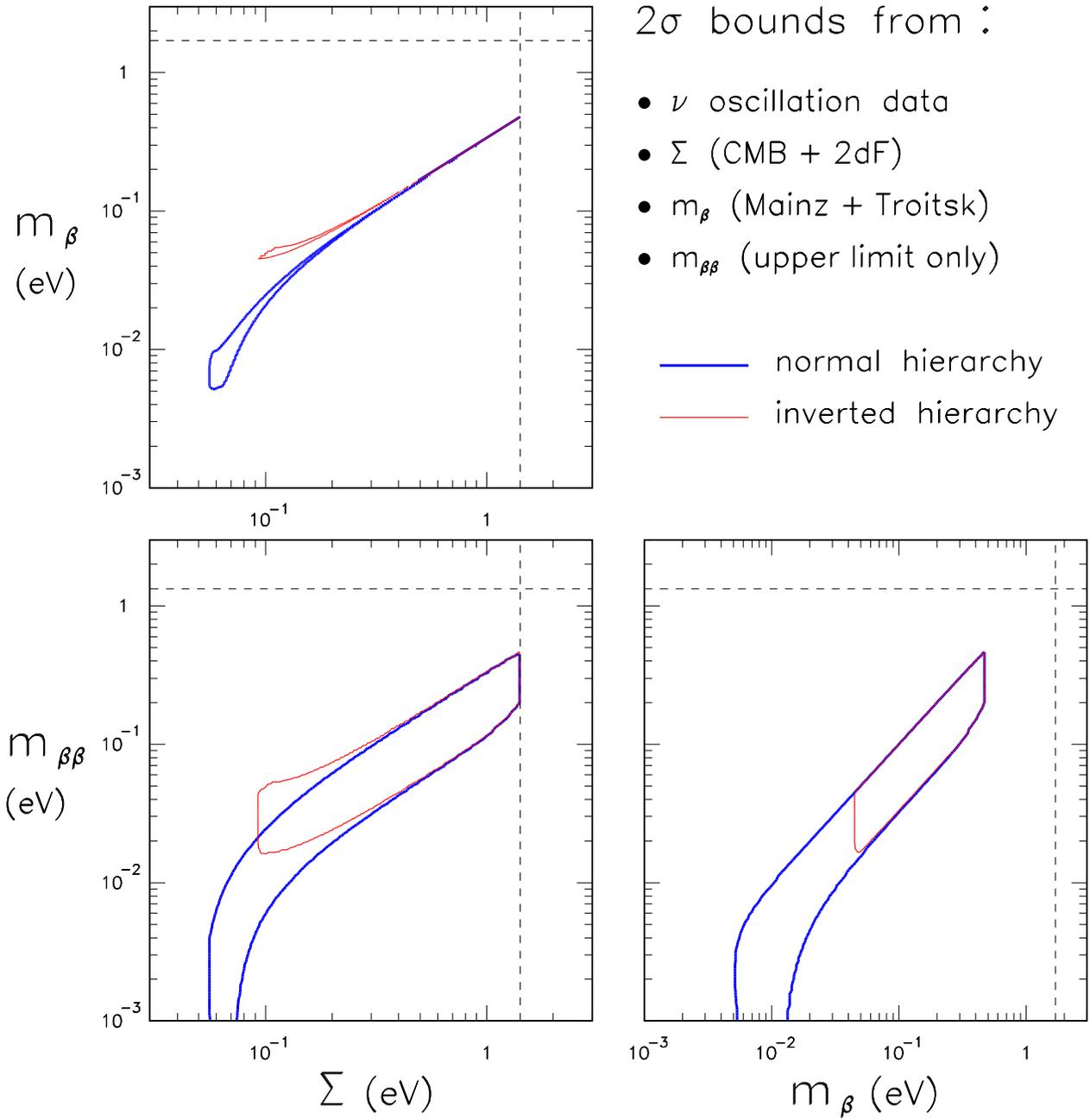}
\vspace*{-1cm} \caption{\label{fig4} Global $3\nu$ analysis in the
$(m_{\beta},m_{\beta\beta},\Sigma)$ parameter space, using
oscillation data plus laboratory data and cosmological  data. With
respect to Fig.~3, this figure implements also upper limits (shown
as dashed lines at $2\sigma$ level) on $m_\beta$ from Mainz+Troitsk
data, on $m_{\beta\beta}$ from $0\nu2\beta$ data, and on $\Sigma$
from CMB+2dF data. In combination with oscillation parameter bounds,
the cosmological upper limit on $\Sigma$ dominates over the
laboratory upper limits on $m_\beta$ and $m_{\beta\beta}$.}
\end{figure}

\begin{figure}
\vspace*{-0cm}\hspace*{-2.5cm}
\includegraphics[scale=0.9, bb= 30 100 500 700]{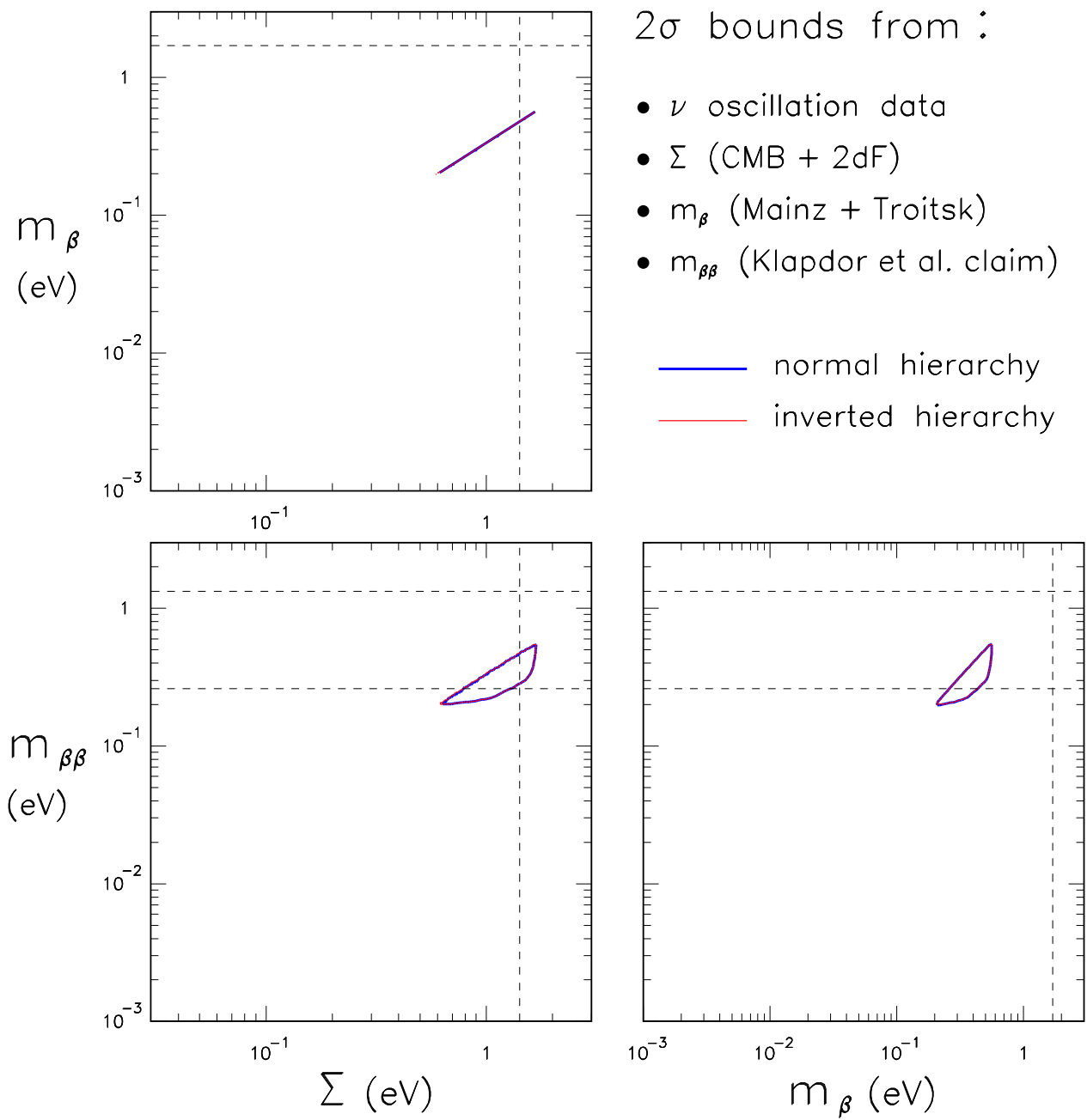}
\vspace*{-1cm} \caption{\label{fig5} Global $3\nu$ analysis in the
$(m_{\beta},m_{\beta\beta},\Sigma)$ parameter space, using
oscillation data plus laboratory and cosmological data. With respect
to Fig.~4, lower bounds on $m_{\beta\beta}$ from a claimed
$0\nu2\beta$ signal are implemented. The globally allowed $2\sigma$
regions appear to be stretched somewhat beyond the separate
$2\sigma$ bands. This feature reflects some tension existing between
the $0\nu2\beta$ claim and cosmological CMB+LSS data within the
$3\nu$ oscillation scenario.}
\end{figure}

\begin{figure}
\vspace*{-0cm}\hspace*{-2.5cm}
\includegraphics[scale=0.9, bb= 30 100 500 700]{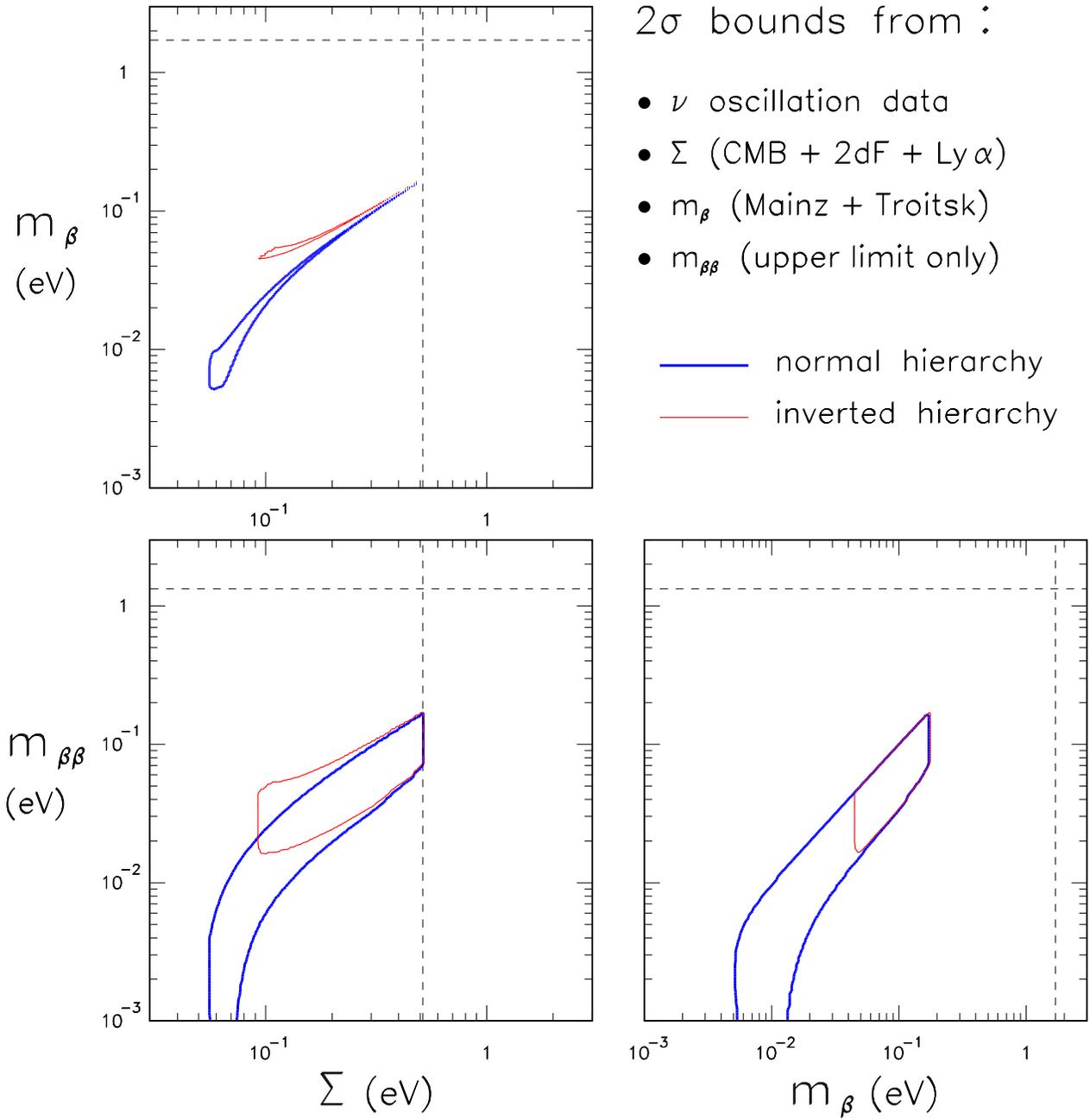}
\vspace*{-1cm} \caption{\label{fig6} Global $3\nu$ analysis in the
$(m_{\beta},m_{\beta\beta},\Sigma)$ parameter space as in Fig.~4,
but including Ly$\alpha$ forest data from the SDSS, which lead to a
stronger upper bound on $\Sigma$, close to the zone where normal and
inverted hierarchies start to branch out in the $(m_\beta,\Sigma)$
and $(m_{\beta\beta},\Sigma)$ planes.}
\end{figure}

\begin{figure}
\vspace*{-0cm}\hspace*{-1.8cm}
\includegraphics[scale=0.85, bb= 30 100 500 700]{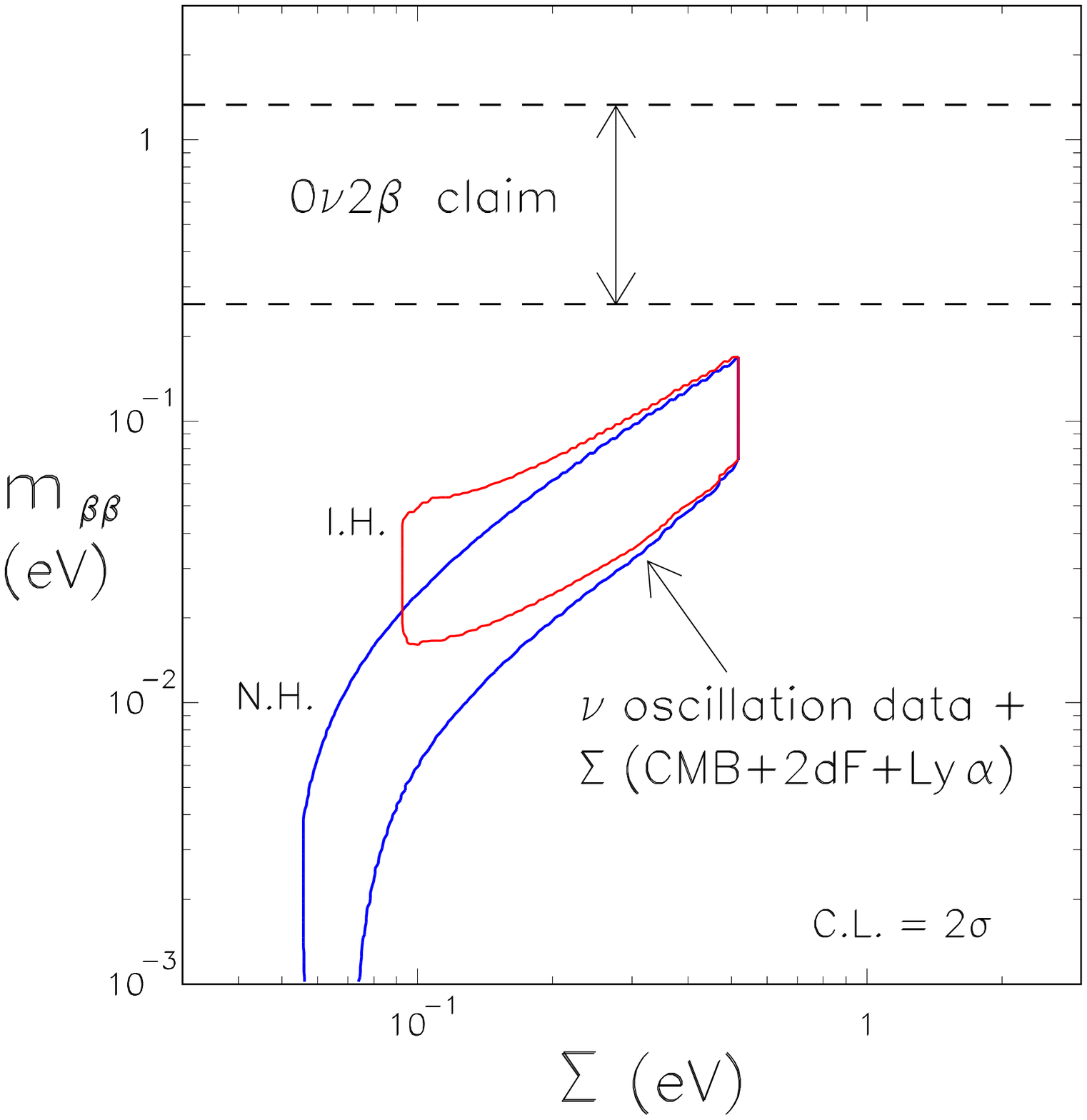}
\vspace*{-1cm} \caption{\label{fig7} $3\nu$ analysis in the
$(m_{\beta\beta},\Sigma)$ plane, showing the absence of overlap
between the regions separately allowed at $2\sigma$ by the
$0\nu2\beta$ claim (horizontal band) and by the combination of $\nu$
oscillation data with CMB+2dF+Ly$\alpha$ cosmological data (slanted
bands), for both normal hierarchy (N.H.) and inverted hierarchy
(I.H.). Such strong tension might indicate possible problems either
in some experimental data or in their theoretical interpretation.}
\end{figure}


\begin{thebibliography}{99}
\bibitem{Dago}  G.~D'Agostini,
                physics/0403086. The homepage
                www-zeus.roma1.infn.it/$^\sim$dagos also contains
                extensive references to statistical
                issues in physics.
\end{thebibliography}
\end{document}